\documentstyle[prd,aps,preprint,epsfig,floats]{revtex}
\tightenlines

\def\DESepsf(#1 width #2){\epsfxsize=#2 \epsfbox{#1}}
\begin{document}

\draft

%@twocolumnfalse\endcsname
\preprint{\vbox{
\hbox{UMD-PP-04-036}
}}

\title{{\Large\bf Neutrino Mass, Dark Matter and Inflation}}
\author{\bf D. Kazanas$^{1}$, R.N. Mohapatra$^2$, S. Nasri$^2$ and
V. L. Teplitz$^{1,3}$ }

\address{$^1$ NASA Goddard Space Flight Center, Greenbelt, MD 20771\\
$^2$ Department of Physics, University of Maryland, College Park,
MD
20742, USA\\
$^3$ Department of Physics, SMU, Dallas, TX 75275}
\date{March, 2004}
\maketitle

\begin{abstract}
We show that spontaneous breaking of global $B-L$ symmetry responsible for
small neutrino masses via the seesaw mechanism provides a unified picture
of hybrid inflation and  dark matter if the scale of $B-L$ breaking is
close
to the GUT scale. The majoron which acquires a small mass in the
milli-eV range due to Planck scale breaking of $B-L$ is the dark matter
candidate. The coupling of the majoron field to the neutrino induces a
small violation of CPT and Lorentz invariance at the present epoch. We
discuss some of the phenomenological and cosmological implications of the
model.
\end{abstract}

\vskip1.0in
\newpage

\section{Introduction}
While a non-zero neutrino mass is at the moment the only experimental 
evidence for physics beyond the standard model, as far as particle physics is 
concerned, in the domain of cosmology, several other oquestions remain 
unanswered within the standard model framework and also cry out for new
physics. They are: ($i$)  particle physics candidates for dark matter,
($ii$) a deeper understanding of the mechanism of inflation
\cite{Lyth:1998xn}, ($iii$) the nature of matter anti-matter asymmetry 
responsible for the observable universe\cite{Riotto:1999yt} as well as ($iv$) 
an explanation of dark energy. 
There are many interesting and compelling models of new physics that can 
explain the different cosmological phenomena listed above individually. 
However, it is always much more desirable that a single model explain 
more than one phenomenon (ideally, of course, all of them). In this paper 
we discuss a model\cite{zhang,cmp} originally designed to explain neutrino 
masses via the seesaw mechanism that seems to provide, after supersymmetrization, 
an explanation of both inflation and dark matter in a rather novel manner. 
The model has other desirable features such as gauge coupling unification, 
stable dark matter, stable proton due to R-parity conservation, etc.

We use  the supersymmetric singlet
majoron \cite{cmp} model described in Ref.\cite{zhang} where the
seesaw mechanism\cite{seesaw} for small neutrino masses is implemented
by breaking a global $B-L$ symmetry. The idea is to extend the
standard model by the addition of one right handed neutrino per family and
 three standard model singlet superfields $S, \Delta$ and
$\overline{\Delta}$. Of the new Higgs fields, $S$ has $B-L=0$, $\Delta$
has $B-L =-2$ and $\overline{\Delta}$ has $B-L=+2$. The theoretical
rationale for $B-L$ symmetry is rooted in the present neutrino oscillation 
results which require that the seesaw scale be much lower than the Planck
scale. It is therefore natural to think that it is protected by some
symmetry. The simplest symmetry that
does this is the $B-L$ symmetry. $B-L$ symmetry could be a global
or a local symmetry. If we take it to be a global symmetry, its
spontaneous breaking  leads to a Nambu-Goldstone boson,
the majoron. Since it is natural to expect that all global
symmetries are broken by nonperturbative gravitational (or
stringy) effects, we will parameterize these symmetry breaking effects by
nonrenormalizable terms in the effective Lagrangian which are suppressed
by the Planck scale $M_{P\ell}$. These Planck scale $B-L$ breaking effects
then give a tiny mass to the majoron\cite{akhm}.

In this paper we show that this model leads to: ($i$) a picture of hybrid
inflation of the universe
that links the neutrino mass (more precisly the seesaw or
$B-L$ scale of order $10^{15}$ GeV) to inflation with all the
desired features and ($ii$) 
Planck scale $B-L$ breaking effects that lead to a mass for the majoron 
in the milli-electron volt range and make it a candidate for dark
matter of the Universe in the same way as the familiar axion, even though
the parameters of the model are very different.  It is, of course,
interesting that this model ties the neutrino
mass and the scale of inflation to the scale of gauge coupling
unification.

An interesting implication of the model is that it leads to a
cosmologically induced Lorentz violation for neutrinos.
 We also find that the superpartners of the
majoron must be in the MeV range in order to be compatible with the
successes of big bang nucleosynthesis (BBN).

\section{The model}
As already noted, the supersymmetric singlet majoron model
\cite{zhang} consists of the following
superfields in addition to the well known quark, lepton, Higgs and gauge
fields of the minimal supersymmetric standard model (MSSM): a right
handed neutrino field $N^c_i$ ($i$ is the generation index), new Higgs
fields $\Delta, \bar{\Delta}$ which carry lepton number $B-L=\pm 2$ and a
singlet field $S$, which is
$B-L$ neutral. We will show below that this model with an appropriate
choice
of the superpotential given below leads to F-term inflation as well as to the
majoron as the dark matter candidate. We require the renormalizable part of the
theory to be
invariant under global $U(1)_{B-L}$ or an effective global symmetry which
contains $B-L$ as a subgroup\footnote{For instance one could have a local
$B-L$
model like a supersymmetric model based on the gauge group
$SU(2)_L\times U(1)_{I_{3R}}\times U(1)_{B-L}$ and have local $B-L$
symmetry
be broken by a multiplet ($\Sigma$) with $B-L \geq 6$. The effective low
energy
theory in this case has a global U(1) symmetry which behaves like
global
$B-L$. One can construct a D-term inflation\cite{Binetruy:1996xj} model 
based on this model with a pair 
of $\Delta$ and $\Sigma$ fields (and their conjugates), if we choose a
form
for the
superpotential 
$W~=~\lambda S(\Delta\overline{\Delta} - M^2_1)+ \lambda_1
X\Sigma\overline{\Sigma}$ and a Fayet-Illiopoulos term for the gauged
$U(1)_{B-L}$. We do not elaborate on this model here. It has all the
properties of the model presented here, with somewhat different
parameters.} .

The superpotential for the model can be written as a sum of three terms:
\begin{eqnarray}
{\bf W}~=~ {\bf W_{MSSM}}~+~{\bf W_0}~+~{\bf W_1}.
\end{eqnarray}
where ${\bf W_{MSSM}}$ is the familiar superpotential for the MSSM; ${\bf
W_0}$ is the renormalizable part involving the new fields of the model and
has the following form:
\begin{eqnarray}
{\bf W_0}~=~h_\nu LH_u N + fNN\Delta+ \lambda S(\Delta \bar{\Delta} -
v^2_{BL}).
\end{eqnarray}
We assume that nonperturbative Planck scale effects  induce
nonrenormalizable terms in the potential and since we expect them to
vanish in the limit of vanishing Newton's constant, they should be
suppressed by powers of $M_{P\ell}$. We arrange our theory such that the
leading Planck scale induced term has the form
\begin{eqnarray}
{\bf W_1}~=~\frac{\lambda_1}{M^3_{P\ell}}(H_uH_d)^2 (\Delta)^2
\end{eqnarray}
At the nonrenormalizable level, there are many terms that one
could write but we find this to provide a good description of
physics of interest here and such a form could be guaranteed
by additional discrete symmetries. For instance a symmetry under
which $S\rightarrow  S; \; \Delta \rightarrow i\Delta; \;
\bar{\Delta}\rightarrow -i\bar{\Delta}; \; H_u\rightarrow H_u; \;
H_d\rightarrow i H_d; \; \nu^c\rightarrow e^{-i\pi/4}\nu^c; \;
L\rightarrow e^{i\pi/4} L; \; e^c\rightarrow e^{5i\pi/4}e^c; \;
d^c\rightarrow -id^c$  allows all the required terms for MSSM
except the $\mu$-term. We include the $\mu$-term in the
superpotential since it is a lower dimensional term and it will break
the symmetry softly.

This superpotential leads to a potential involving the fields $S,
\Delta$ and $\bar{\Delta}$ with the following form:
\begin{eqnarray}
V(\Delta, \bar{\Delta}, S)~=~\lambda^2|\Delta\bar{\Delta} - v^2_{BL}|^2+
\lambda^2|S|^2(|\Delta|^2+|\bar{\Delta}|^2) ~+~V_{SSB};
\end{eqnarray}
where $V_{SSB}$ stands for the supersymmetry breaking terms such as
$m^2_S|S|^2$ and $|m^2_1|\Delta|^2+m^2_2|\bar{\Delta}|^2$ etc.
 where these supersymmetry breaking mass parameters are all in the TeV
range. We have ignored the Planck scale $B-L$ breaking terms since they
are
small and not relevant to the discussion of inflation given in the next
section. The minimum of this
potential corresponds to $\langle \Delta \rangle = \langle 
\bar{\Delta} \rangle = v_{BL}$.

Once the $B-L$ symmetry is broken, this leads to the seesaw mechanism for
small neutrino masses. Typically, a neutrino mass has the form
$m_\nu\simeq \frac{m^2_{\nu^D}}{M_R}$ where $m_{\nu^D}$ is of the same
order as a typical fermion mass of the standard model and
$M_R=fv_{BL}$ with $f \simeq 1$. If we assume that
the atmospheric neutrino oscillation is linked to the third generation
quarks (as would it be plausible in a quark-lepton symmetric theory), then
the Dirac mass would be about 100 GeV and this would give
$v_{BL}\simeq 10^{15}$ GeV, which is close to the value preferred by the
inflation picture discussed below.

 Because the vevs of $\Delta$ and $\bar{\Delta}$ break the global $B-L$
symmetry, in the absence of explicit $B-L$ breaking terms
(denoted by $V_{Planck}~=~0$), we have a massless particle in the theory, 
the majoron,  given by $\phi\equiv (\chi-\bar{\chi})/\sqrt{2}$, where
we have parameterized $\Delta =\frac{1}{\sqrt{2}}
(v_{BL}+\rho)e^{i\chi/v_{BL}}$ and  $\bar{\Delta} =\frac{1}{\sqrt{2}}
(v_{BL}+\bar{\rho})e^{i\bar{\chi}/v_{BL}}$.
The potential for $\phi$ is flat due to the shift symmetry under
which $\phi \rightarrow \phi +\alpha$. Once the Planck scale terms are
turned on, the potential $V(\phi)$ loses its flatness and can generate
a rolling behaviour for the $\phi$ field as we see below.

\subsection{Inflation}
This model has the ability to generate inflation in early stages
of the Universe. This comes about due to the interplay among the
fields $S$ and $\Delta$ and $\bar{\Delta}$. The potential
involving them is given by Eq.(2), which has the form required in
the hybrid inflation scenario. To see this note that for, $S\geq
v_{BL}$, the minimum of the potential corresponds to
$\Delta=\bar{\Delta}=0$. We assume that $\lambda
\left(\frac{v_{BL}}{M_{P\ell}}\right) > (\frac{m_S}{v_{BL}})$. The
potential is
then dominated by the term $V_0=~ \lambda^2 v^4_{BL}$ and the
universe undergoes an inflationary phase. The $S$ field 
keeps rolling towards the potential minimum and inflation ends when 
the field $S$ reaches the value $S\simeq v_{BL}$. However it is 
necessary for the potential
to have small tilt to drive the field toward its global minimum.
This is the hybrid inflation picture\cite{linde}.
%\subsection{$\lambda \geq 10^{-4}$}

One way to generate a slope along the inflationary trajectory (i.e. the
S direction) is to include the one loop radiative correction to
the tree level potential \cite{DSS}. This arises because
supersymmetry is broken during inflation and there will be a mass
splitting between the components in the chiral  multiplet. The
effective one loop radiative correction is
\begin{equation}
\Delta V =\sum_i \frac{1}{64\pi^2}(-1)^{2J_i}(2J_i +
1)\textit{M}^4_i(S)\ln{\left(\frac{\textit{M}_i^2(S)}{\Lambda^2_R}\right)}
\end{equation}
Where the sum extends over all the spin states $J_i$ with field
dependent mass $\textit{M}_i$ and $\Lambda_R$ is a renormalization
scale. The $F_S\equiv \frac{\partial W}{\partial S} \neq 0$ term splits
the components of the chiral
multiplets $\Delta$ and $\bar{\Delta}$ into a pair of two real
scalar fields with masses $m^2_{\pm} = \frac{\lambda^2}{2}(S^2 \pm
2v^2_{BL})$ and a Dirac fermion with mass squared $m^2_F =
\frac{\lambda^2}{2}S^2$. Since during the inflation $\Delta =
\overline{\Delta}=0$, the effective potential reads
\begin{equation}
V_{eff} $=$ \lambda^2 \{v_{BL}^4 + \frac{\lambda^2}{128\pi^2}[(S^2
-2v_{BL}^2)^2\ln{(\frac{\lambda^2(S^2 - 2v_{BL}^2)}{\Lambda^2})} +
(S^2 + 2v_{BL}^2)^2\ln{(\frac{\lambda^2(S^2 +
2v_{BL}^2)}{\Lambda^2})}
-2S^4\ln{(\frac{\lambda^2S^2}{\Lambda^2})}] \}
\end{equation}
For $S$ much larger than the $(B -L)$ symmetry breaking scale the
above potential becomes
\begin{equation}
V_{eff} \simeq \lambda^2 v_{BL}^4\{1 +
\frac{\lambda^2}{16\pi^2}\ln{\frac{S}{v_{BL}}}\}
\end{equation}
The log term will provide the driving force for the field $S$ to
roll down the potential. In this case the slow roll parameters are
given by
\begin{equation}
|\eta| = \frac{1}{2N(S)} \gg \epsilon
=\frac{\lambda^2}{32\pi^2}\frac{1}{N(S)}
\end{equation}
where $N$ is the number of e-foldings and is given by
\begin{equation}
N(S) = \frac{32\pi^3}{\lambda^2 M^2_{P\ell}}S^2
\end{equation}
Density perturbations in about $N_{60} = 60$ e-foldings
before the end of inflation are estimated as
\begin{equation}
\frac{\delta\rho}{\rho} \simeq
16\pi\sqrt{\frac{N_{60}}{3}}(\frac{v_{BL}}{M_{P\ell}})^2
\end{equation}
%\begin{equation}
% S_{end} = max \{v_{BL}, S(\epsilon =1), S(\eta
%$\end{equation} where $\epsilon$ and $\eta$ are the slow
%roll parameters and given by
Fluctuations with amplitude $\sim 10^{-5}$ can obtained if $v_{BL}
\simeq 6 \cdot 10^{15} \;GeV$.

Another way to generate a slope to the potential is to add soft
supersymmetry breaking mass $m_S$ to the inflaton field \cite{C}.
In this case the slow roll parameters are given by
\begin{eqnarray}
\epsilon = 4\pi\frac{M_{P\ell}^2m_S^2}{V_0}(\frac{m_S^2S^2}{V_0})\nonumber \\
\eta = 8\pi\frac{M_{P\ell}^2m_S^2}{V_0}\gg \epsilon
\end{eqnarray}
The requirement that $\eta\ll 1$ (say $0.01$) gives
\begin{equation}
\lambda \simeq
10^{3/2}(\frac{M_{P\ell}}{v_{BL}})(\frac{m_S}{v_{BL}})
\end{equation}
In order for the soft supersymmetry breaking to dominate over the
one loop radiative correction the parameter $\lambda$ must be
smaller than $3 \cdot 10^{-5}$ . The value of $S$ after suffering
$N_{60}\simeq 60$ e-foldings between the horizon exit and the the
end of inflation is given by
\begin{equation}
S_{60} = v_{BL}\exp{(60\eta)}
\end{equation}
The $\Delta$ field plays the role of the inflaton field, which then
oscillates and reheats the Universe via the production of
neutrinos due to $NN\Delta$ coupling. The density fluctuations
lead to strong constraints on the parameters of the model and 
are given by :
\begin{eqnarray}
\frac{\delta \rho}{\rho}\simeq 
\lambda^3\sqrt{8\pi}(\frac{v_{BL}}{M_{P\ell}})^3
(\frac{v_{BL}}{m_S})^2\exp{(-60\eta)};
\end{eqnarray}
For $v_{BL}\simeq 2 \cdot 10^{14}$ GeV , $m_S\simeq 2$ TeV and
$\lambda\sim 10^{-{4.5}}$ we get $\delta \rho/\rho\simeq 10^{-5}$.
It is interesting that the value for $v_{BL}$ required for
understanding neutrino masses also gives the correct order of
magnitude for density fluctuations generated by inflation.

Finally, we discuss the question of reheating after inflation ends.
The reheating can be assumed to be caused by the decay of the field
$S$ when it starts oscillating around the minimum. The reheating temperature
is given by the approximate formula $T_R\sim \sqrt{M_{P\ell} \Gamma_S}$
where $\Gamma_S$ is the decay width of the inflaton. To obtain the decay
width of $S$, we have to isolate the decay modes. We expect that the
$\lambda$ coupling in the renormalizable part of the superpotential will
give the dominant contribution to the decay width.

To see this in detail, note that if both $\Delta$ and
$\overline{\Delta}$ have equal vev\footnote{these vevs will
differ by a small amount proportional to the supersymmetry
breaking which can therefore be neglected for large values of
$v_{BL}$ that we are interested in.}, the combination $\psi\equiv
\frac{1}{\sqrt{2}}(\psi_\Delta -\psi_{\bar{\Delta}})$ is the
fermionic partner of the majoron  field (denoted by $\phi$) and is
therefore light
whereas
$\frac{1}{\sqrt{2}}(\psi_\Delta + \psi_{\bar{\Delta}})$ is the
fermionic partner of the superheavy supermultiplet and therefore
has same mass as $S$ field. However one can see from the
superpotential that $S$ has a coupling to $\psi\psi$ and can
therefore decay to these states.  The decay width is then given by
$\Gamma_S\sim \frac{\lambda^2}{4\pi} m_S\simeq 10^{-14.5}v_{BL}$.
Note that the inflaton gets its dominant mass from the $B-L$ breaking vev
$v_{BL}$ and a smaller mass from supersymmetry breaking.Therefore
$M_S\simeq v_{BL}$.
This leads to a reheating temperature of $T_R \simeq 10^{9}$ GeV. In
this case a gravitino with mass
$m_{3/2}\simeq 100 GeV$ does not spoil the success of big bang
nucleosynthesis.

It is also worth pointing out that the SUGRA embedding of this model os
free of the $\eta$ problem that is generic to F-term SUGRA
models\cite{Binetruy:1996xj}. 

\subsection{Nonrenormalizable terms, majoron mass and majoron as the
ultralight dark matter:}
In the presence of the nonrenormalizable terms, the majoron picks up a
nonzero mass. If we choose the leading order non-renormalizable term to be
$\frac{(H_uH_d)^2(\Delta ^2)}{M^3_{P\ell}}$ (which can be done by a
judicious choice of discrete symmetries as discussed above), then we get
the majoron
mass $m^2_\phi\simeq \frac{\mu v^4_{wk}}{M^3_{P\ell}}\simeq
(10^{-15}~eV)^2$.

In order to discuss how massive majorons constitute the dark matter of
the universe, let us discuss the evolution of the majoron field
$\phi$ as the universe evolves. The majoron potential, which arises solely
from the $B-L$ breaking terms can be written using Eq. (3) in the form:
\begin{eqnarray}
V(\phi)~=\lambda_1\Lambda^4(1+ cos \frac{\phi}{v_{BL}}) ~, 
~~~~\lambda_1 \simeq 1
\end{eqnarray}
where $\lambda_1\Lambda^4 = m^2_\phi v^2_{BL}\simeq (1.7
\times 10^{-3}~GeV)^4$. We have added a constant term, akin to a 
cosmological constant, so that after the amplitude for the $\phi$ field
oscillation damps to its minimum at $\phi = \pi v_{BL}$ the value of the 
potential is zero. The field $\phi$ and the radiation energy density
$\rho$ satisfy the coupled scalar field - Einstein-Friedman equations
\begin{eqnarray}
\ddot{\phi} + 3H\dot\phi = - V^{\prime}(\phi)
\label{phi}
\end{eqnarray}
and 
\begin{eqnarray}
H^2~=~\frac{8\pi G_N}{3}[\rho + \frac{1}{2}\dot{\phi}^2+ V(\phi)]
\label{hubble}
\end{eqnarray}
In the early universe, clearly the $V^{\prime}$ term is
negligible. The Hubble parameter $H$ is then dominated either by
the kinetic energy of $\phi$ or the relativistic energy density in
radiation. Solving Eq. \ref{phi}, one finds that $\dot\phi\propto R^{-3}$
where $R$ is the scale factor. So regardless of whatever initial
$\dot\phi$ the field starts out with, it completely damps down to
a very small value as the universe expands. Thus the value of
$\phi$ freezes over for most of the early universe at some random
value. When $m_\phi\simeq 3 H$, the potential term will dominate Eq.
(16) and the field will oscillate around its minimum value of
$\phi=\pi v_{BL}$. These oscillations, which do not damp, behave like 
matter \cite{turner} and contribute to the dark matter
density. Below, we calculate this contribution and find that for
the range of parameters of interest, it gives the right order of
magnitude for $\Omega_{matter}$.

As noted already, the value of the $\phi$ field remains frozen at
its initial value (taken to be $v_{BL}$) until the epoch when
$m_\phi\simeq 3H\simeq \frac{3T^2_i}{M_{P\ell}}$ at which time the
$\phi$ field starts to oscillate. The temperature at which it does
this is given by $T_i\simeq \sqrt{M_{P\ell}m_\phi/3}\simeq 10^{-3}$
GeV. As it oscillates, it is easy to show that it acts like a
pressureless gas and thus can be treated as an ensemble of
nonrelativistic particles\cite{turner}. The contribution of the
majoron to the energy density now is given by: 
\begin{eqnarray}
\Omega_\phi\simeq\frac{ m^2_\phi
v^2_{BL}}{H^2_0M^2_{P\ell}}\left(\frac{T_0}{T_i}\right)^3\simeq 0.1.
\end{eqnarray}
The ratio of the majoron (dark matter)  to the radiation energy density 
at their equipartition temperature $T_{EQ} \sim 1$ eV is given by
\begin{equation}
\frac{\rho_\phi}{\rho_R}(T_{EQ})\simeq \frac{\Lambda^4
T_0^4}{\rho^{(0)}_R T_{EQ} T^3_i}
\end{equation}
where $T_0 \simeq 2.4 \cdot 10^{-4}$ eV and $\rho^{(0)}_R \sim
2 \cdot 10^{-15}$ eV$^4$ are the temperature and the energy density of
the photons today. For $\lambda_1\Lambda^4 \simeq 1$ MeV$^4$ and 
$T_i \sim 100$ MeV one obtains $\frac{\rho_\phi}{\rho_R}(T_{EQ}) 
\sim 1$. Hence the universe remains radiation dominated until the 
temperature drops below $\simeq 1$ eV.

Thus we see that majoron oscillations can act as a nonrelativistic
dark matter with the right order of magnitude for
$\Omega_{matter}$. We therefore conclude that the same range of
parameters that gave the neutrino mass values in the right order
and also the density fluctuations required by CMB measurements
also seems enable the majoron to play the role of dark matter.

To make these ideas concrete, we have solved numerically the 
coupled equations (\ref{phi}) and (\ref{hubble}) in the regime prior to 
and after the majoron field begins oscillating. That is the region in
which the Hubble parameter, which includes the sum of radiation, matter and
the majoron field energies, in the second term in the right hand side of
Eq. (\ref{phi}), has decreased to the point that the right hand side has become
appreciable in comparision with that term. For natural choices of the
parameters discussed above ($v_{BL}$ and $v_{WK}$ $10^{-4}M_{P\ell}$ and 100
GeV respectively) the coefficient $\Lambda^4$ in the Eq. (15) becomes
$(10^{-4}~{\rm GeV})^4$. In Figures 1 and 2, we plot the solution for 
the majoron field
$\phi$ for two initial $\phi$ values: $\phi/v_{BL} = 0.1\pi$ and $ \pi/3$.
We note that the amplitude of the final field oscillation is less in the
second case. For still larger initial values of $\phi$, the amplitude
falls off more sharply. On the other hand, both for smaller initial $\phi$
values and over a wide range of $\dot\phi$ values, the final result for
the majoron oscillation amplitudes are the same.

In Figures 3 and 4 we plot the resulting ratio $\rho_\phi/\rho_\gamma$ 
for the same choice of parameters as in Figures 1 and 2 respectively, over a
broad region in which the oscillatory frequency is not too large compared 
to the expansion rate so that numerical computations with reasonable number 
of steps can be carried out and are reliable. One sees that this ratio 
increases linearly with decreasing temperature, as is appropriate for 
presureless matter. One can find today's value for $\rho_\phi/\rho_\gamma$ 
simply by multiplying the values read-off the figure by $T/T_0$. These numerical computations confirm that our parameter choice
leads to the correct order of magnitude for today's ratio (that is,
oscillating $\phi$ field giving $\Omega_{DM}\sim 0.3$). 

Finally we note that the frequency of the majoron field oscillation is
given by $\omega = \left(\lambda_1\Lambda^4/v^2_{BL}\right)^{1/2}\simeq
0.2 \; {\rm sec}^{-1}$. The oscillations begin around the epoch of 
nucleosynthesis. The energy in these oscillations around this epoch is
very small compared to that in $\rho_\gamma$ so that it does not affect 
the BBN considerations. Furthermore it is worth noting that
the frequency $\omega$ is independent of the value of $v_{BL}$.

 \section{Other consequences of the model}
 Some consequences of the superlight majoron for cosmology has
recently been discussed in \cite{chacko}. We consider other implications
that particularly relate to the model in this paper in this section.

\subsection{Cosmologically induced Lorentz violation for neutrinos}
 One important implication of our proposal is that it leads to a
cosmologically induced violation of Lorentz invariance. This comes about
because in our model the majoron field has a derivative coupling of the
form $\frac{1}{v_{BL}}\bar{\nu}\gamma_\mu \nu \partial^{\mu}\phi$.
In the late universe when $\dot{\phi}\neq 0$, this leads to an effective
Lorentz violating term of the form
$\frac{\dot{\phi}}{v_{BL}}\nu^{\dagger}\nu$ in the effective low energy
Lagrangian. This effect is Lorentz violating and will manifest itself in
the neutrino oscillation process\cite{barger}. The maximum value of the
parameter $ \frac{\dot{\phi}}{v_{BL}}$ in our model is $m_\phi\simeq
10^{-23}-10^{-24}$ GeV.  However since the Lorentz
violating term is family universal, it will most likely manifest irself in
a transition from $\nu_\alpha$ to $\bar{\nu}_\beta$. 
 So the only way to detect this will be to measure
$P(\nu_\alpha\to \bar{\nu}_\beta)-P(\bar{\nu}_\beta\to
{\nu}_\alpha)$. In the Lorentz invariant case, such effects are generally
suppressed by the mass of the neutrino\cite{andre} and are therefore
likely to be very small. The detailed experimental
implications of such Lorentz violating effects are currently under
investigation.

\subsection{BBN constraints on majoron superpartners}
 A second phanomenological implication is that if we ignore the
supersymmetry breaking effects, then majoron
belongs to a supermultiplet together with a scalar partner (to be
called smajoron) and a fermion, majorino, both of which are massless. In
the presence of supersymmetry breaking effects, the smajoron
($\sigma$)
and majorino ($\psi$)
pick up mass in the MeV to GeV range. The precise values of these masses
are however constrained by cosmological consideration for the model to
be viable. We discuss them in this section.

This question was discussed in \cite{zhang} for the case where the $B-L$
breaking scale is in the TeV range. It was found that in that case the
masses of $\sigma$ and $\psi$ can be in the TeV range. This holds
as long as $v_{BL} \leq 10^8$ GeV. In our case however $v_{BL} \simeq
10^{15}$ GeV. We will therefore end up with the smajoron ($\sigma$)
and majorino ($\psi$) masses in the 10 MeV or lower range.

The primary goal is to make sure that the new particles do not affect
big bang nucleosynthesis (BBN). Since the couplings of the $\phi,
\sigma$
and $\psi$ to standard model particles are all suppressed by $v_{BL}$,
the lifetimes of $ \sigma$ and $\psi$ are much longer than one second,
the BBN epoch. Therefore if they are heavy and their abundances at the BBN
epoch are not suppressed, then they will have adverse effect on
nucleosynthesis. We therefore have to calculate their abundance at the
BBN epoch.

Due to the suppressed  couplings of $\phi, \sigma$
and $\psi$, they will decouple in the
very early stage of the universe. To calculate
the decoupling temperature, we note that the typical annihilation rates are
for $\phi, \sigma$ and $\psi$ are given by $R\simeq
\frac{T^5}{v^4_{BL}}$. Using the decoupling condition $R(T_D)  < H(T_D)$,
we get
\begin{eqnarray}
T^3_D~\simeq ~ g^{1/2}_*(T_D) \frac{v^4_{BL}}{M_{P\ell}}
\end{eqnarray}
which leads to $T_D\simeq 10^{-1} v_{BL}$ which is of order $10^{14}$ GeV.
After decoupling the $\phi, \sigma$ and $\psi$ densities simply dilute
due to the expansion of the Universe. Their contribution to the energy
density of the universe at the BBN epoch is given by
\begin{eqnarray}
\frac{\rho_{\phi, \sigma,\psi}}{\rho_\gamma}\simeq \frac{n_{\phi,
\sigma,\psi}}{n_\gamma} \frac{m_{\phi, \sigma,
\psi}}{T_{BBN}}\simeq \frac{g_*(T_{BBN})}{g_*(T_D)}\frac{m_{\phi,
\sigma,\psi}}{T_{BBN}}
\end{eqnarray}
Success of BBN requires that $\frac{\rho_{\phi,
\sigma,\psi}}{\rho_\gamma}$ should be much less than one.
Since $\frac{g_*(T_{BBN})}{g_*(T_D)}\simeq 100$, we therefore must
have $m_{ \sigma,\psi} \ll 100$ MeV.

We further note that since these particles $\phi, \sigma, \psi$ are
singlets under the standard model, they are not
in conflict with any known low energy observations despite their small
mass.

\subsection{Implications for leptogenesis}
Finally, we wish to comment that in this model one can employ
the mechanism of leptogenesis\cite{fuku} to understand the origin of
matter in the universe. One can have a right handed neutrino\cite{buch}
at the intermediate scale range of $10^{9}$ GeV range (or even a pair of
them nearly degenerate\cite{Akhmedov:2003dg}), whose decay would produce
a lepton asymmetry, which via the sphaleron interactions would get
converted to baryons. The only new aspect of our model is the presence of
the majoron at very low energies. In principle its interactions can erase
the baryon asymmetry since its interactions violate lepton number. However
in our case since the scale $v_{BL}$ is very
high, as noted already, the majorons decouple around $10^{14}$ GeV and
are therefore ``impotent'' as far as their effect on lepton asymmetry is
concerned. If for instance the $v_{BL}$ scale was in the range below
$10^{9}$ GeV, we would have no leftover lepton asymmetry at the weak scale
to be converted to baryons. It is therefore interesting that the high
scale of $v_{BL}$ is required from various considerations.

In conclusion, we have presented a simple model for neutrinos using the
supersymmetric extension of the singlet majoron model that provides a
unified framework for understanding inflation and dark matter for the same
set of parameters required by neutrino masses.
 We find that the $v_{BL}$ scale is constrained from various
considerations to be around the conventional grand unification scale $\sim
10^{15}$ GeV.  We have checked our
results using a numerical solution of the evolution equation for the
majoron field. The high $v_{BL}$ scale makes the majoron and its
superpartners highly invisible in collider experiments.

The works of R. N. M. and S. N. are supported by the National Science
Foundation Grant No. PHY-0099544. The works of D. K. and V. L. T. are
supported by NASA. R. N. M. would like to thank Z. Chacko for discussions.

\newpage

\begin{figure}[h!]
\begin{center}
\epsfxsize15cm\epsffile{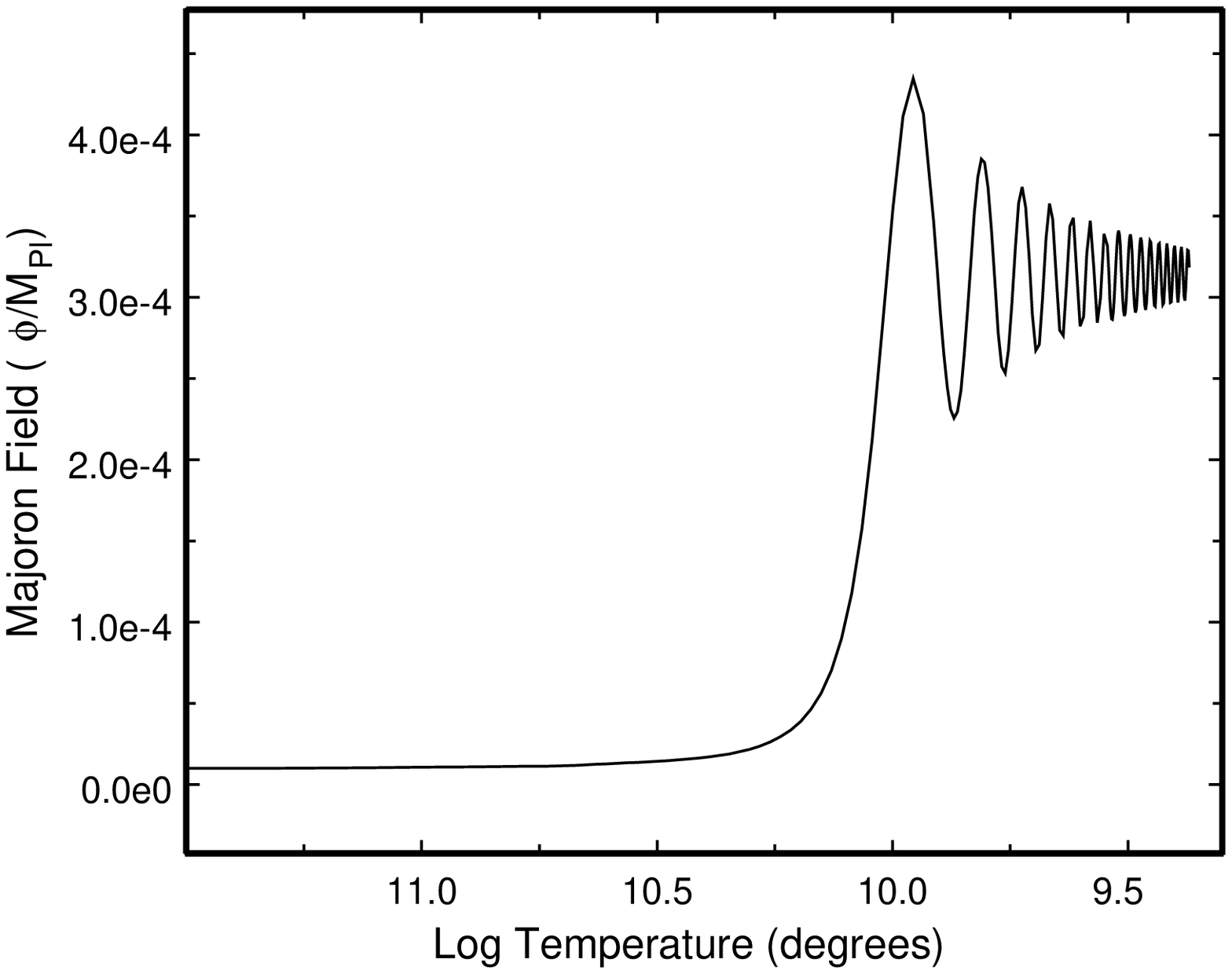}
\caption{ The evolution of the Majoron field (in units of $M_{Pl}$) as a
function of temperature for the initial value of the field $\phi(0) =
10^{-5}$. The field settles down to the value
$\phi = \pi \, v_{BL}; \; v_{BL}= 10^{-4} M_{Pl}$, driven by the first
minumum of the potential. The Log is with respect to base 10 and $e^{-4}$
in the ordinate stands for $10^{-4}$.
\label{fig1}}
\end{center}
\end{figure}

\begin{figure}[h!]
\begin{center}
\epsfxsize15cm\epsffile{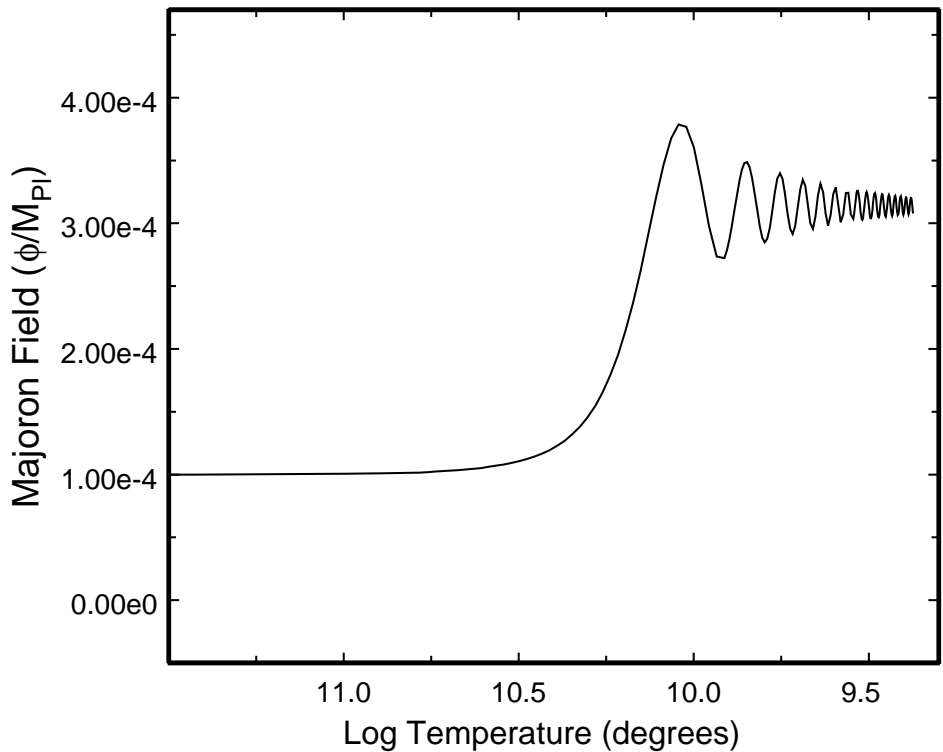}
\caption{The evolution of the Majoron field (in units of $M_{Pl}$) as a
function of temperature for the initial value of the field
$\phi(0) = 10^{-4}$. In this case also, the field settles down to
the same value
$\phi = \pi \, v_{BL}; \; v_{BL}= 10^{-4} M_{Pl}$, driven by the first
minumum of the potential. The notation is same as in Figure 1. Note that
the amplitude of oscillation is smaller than that in Fig. 1 by about a
factor of half.
 \label{fig:cstr1}}
\end{center}
\end{figure}

\begin{figure}[h!]
\begin{center}
\epsfxsize15cm\epsffile{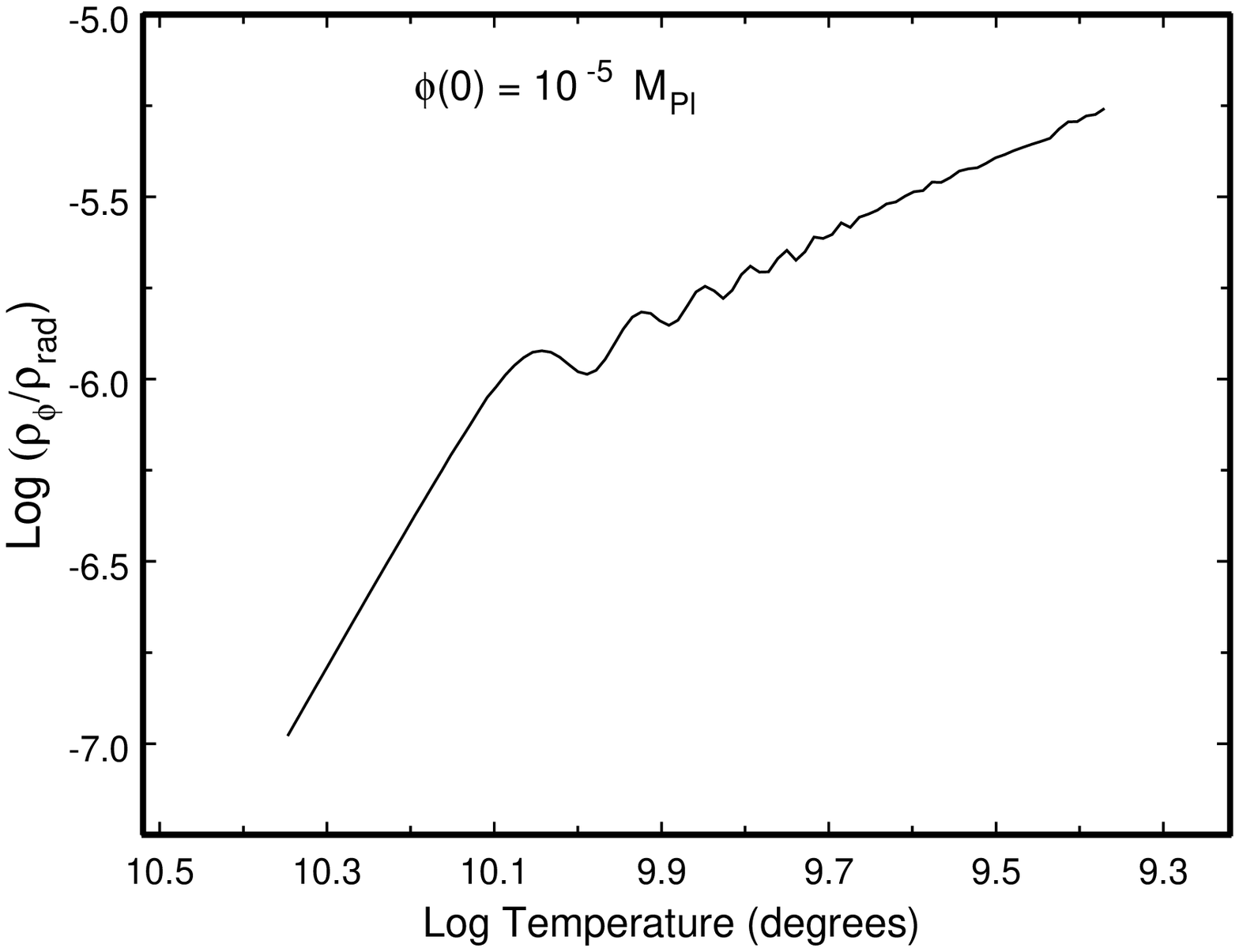}
\caption{The ratio of energy density in the field $\phi$, $\rho_{\phi} =
\dot \phi ^2/2 +V(\phi)$ over the radiation energy density $\rho_{rad}$
as a function of the temperature of the Universe  for initial value
of $\phi(0) = 10^{-5}M_{Pl}$. After the field begins
oscillating it behaves just like matter, as it can be seen from the above
ratio which increases like $T$.
\label{fig3}}
\end{center}
\end{figure}

\begin{figure}[h!]
\begin{center}
\epsfxsize15cm\epsffile{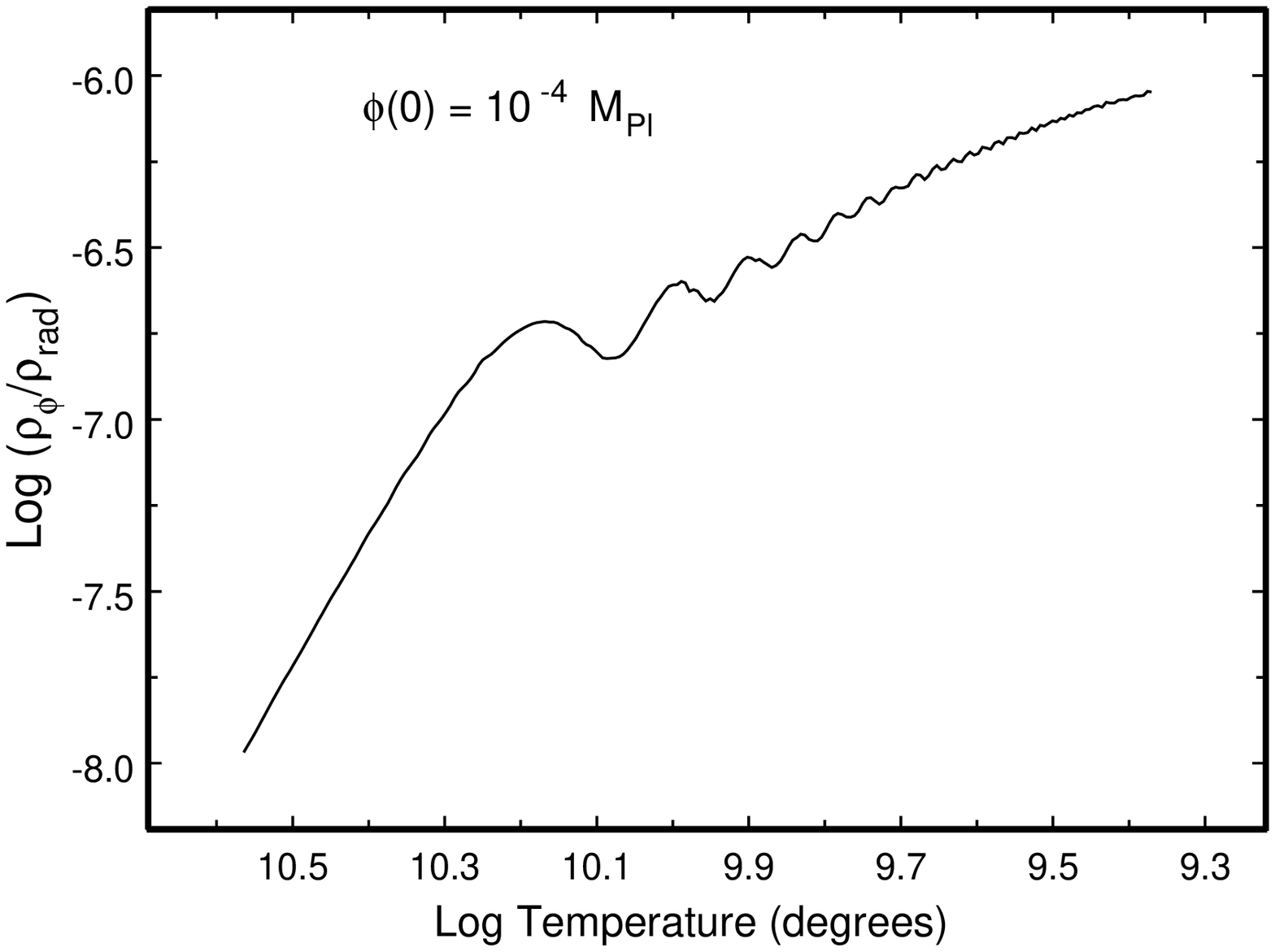}
\caption{
The ratio of energy density in the field $\phi$, $\rho_{\phi} =
\dot \phi ^2/2 +V(\phi)$ over the radiation energy density $\rho_{rad}$
as a function of the temperature of the Universe for initial value
of $\phi(0) = 10^{-4}M_{Pl}$. After the field begins
oscillating it behaves just like matter, as it can be seen from the above
ratio which increases like $T$.
\label{fig4}}
\end{center}
\end{figure}

\end{document}